# Water Agglomerates on Fe$_3$O$_4$(001)

Matthias Meier,[1,2] Jan Hulva,[1] Zdeněk Jakub,[1] Jiří Pavelec,[1] Martin Setvin,[1] Roland Bliem[1], Michael Schmid,[1] Ulrike Diebold,[1] Cesare Franchini,[2] and Gareth S. Parkinson[1]

[1]Institute of Applied Physics, Technische Universität Wien, Vienna, Austria

[2]University of Vienna, Faculty of Physics and Center for Computational Materials Science, Vienna, Austria

**Determining the structure of water adsorbed on solid surfaces is a notoriously difficult task, and pushes the limits of experimental and theoretical techniques. Here, we follow the evolution of water agglomerates on Fe$_3$O$_4$(001); a complex mineral surface relevant in both modern technology and the natural environment. Strong OH-H$_2$O bonds drive the formation of partially-dissociated water dimers at low coverage, but a surface reconstruction restricts the density of such species to one per unit cell. The dimers act as an anchor for further water molecules as the coverage increases, leading first to partially-dissociated water trimers, and then to a ring-like, hydrogen-bonded network that covers the entire surface. Unraveling this complexity requires the concerted application of several state-of-the-art methods. Quantitative temperature programmed desorption (TPD) reveals the coverage of stable structures, monochromatic x-ray photoelectron spectroscopy (XPS) shows the extent of partial dissociation, and non-contact Atomic Force Microscopy (AFM) using a CO-functionalized tip provides a direct view of the agglomerate structure. Together, these data provide a stringent test of the minimum energy configurations determined via a van der Waals density functional theory (DFT)-based genetic search.**

The ubiquity of water in the ambient environment ensures that its interaction with solid surfaces is of fundamental importance [1]. To understand processes such as dissolution, corrosion, and weathering at the molecular level requires an understanding of how water adsorbs on surfaces, and what governs their reactivity. Atomic-scale investigations on single-crystal samples have revealed that interfacial water almost never forms an ice-like structure [2], and aims to simultaneously maximize its interaction with the surface and intermolecular hydrogen bonding (H-bond). The surface and H- bonds have similar magnitude on metals, and the adlayer is stabilized if some fraction of the water dissociates, allowing the formation of strong H$_2$O-OH H-bonds [2].

The situation is somewhat different on metal oxides because the bonds to the surface dominate. The lone pair on the oxygen atom forms a dative bond with the electron-deficient cation sites, while on more reactive surfaces, dissociation gives rise to two distinct hydroxyl groups (terminal O$_{water}$H and surface O$_{surface}$H). The energetic difference between molecular and dissociative adsorption can be extremely small, and some mixture is inevitably observed in equilibrium at finite temperatures [3]. There is, however, increasing evidence that partially-dissociated adlayers can also represent the lowest-energy configuration on metal oxide surfaces [4–6], and partially-dissociated water dimers have been recently proposed to be the most stable water species on both RuO$_2$ and Fe$_3$O$_4$(111) [7–9].

A key issue for the understanding of water adlayers has been the difficulty of achieving molecular resolution of water clusters and adlayers. While significant progress has been made using STM in recent years [10], nc-AFM has emerged as a technique capable of superior resolution, particularly when the tip is functionalized by a CO molecule [11]. Recently, Shiotari and Sugimoto [12] demonstrated spectacular images of water clusters adsorbed on Cu(110), and we resolved to apply this method to the particularly complex case of water adsorption on the ($\sqrt{2}\times\sqrt{2}$)R45°-reconstructed $Fe_3O_4$(001) surface. In combination with quantitative TPD, high-resolution XPS, and state of the art theory, we are able to determine the evolution of stable water structures over the full range from an isolated molecule to the completion of the first monolayer.

Interestingly, although an isolated molecule adsorbs intact, significant energy is gained through the formation of partially-dissociated water dimers. The surface reconstruction limits the coverage of such species to one per ($\sqrt{2}\times\sqrt{2}$)R45° unit cell, however, because only a subset of the surface O atoms can accept a proton to form an $O_{surface}$H group. The partially-dissociated water dimers act as an anchor for further water as the coverage increases, leading first to partially-dissociated water timers, and then to a ring-like H-bonded network, which covers the entire surface. Interestingly, the nc-AFM images allow us to rule out one of two iso-energetic water trimers predicted by a thorough DFT search, and the data indicate that van der Waals DFT does not accurately handle the cooperative energy balance in this system.

**Results**

A key feature of the spectroscopic measurements described here is the ability to deposit an accurately determined number of water molecules on the $Fe_3O_4$(001) surface using a calibrated molecular-beam source [13]. Figure 1(A) shows TPD spectra obtained for various initial $D_2O$ coverages ranging from 0 to 14 molecules per ($\sqrt{2}\times\sqrt{2}$)R45° unit cell ($H_2O$/u.c.). $D_2O$ was utilized to ensure that the measured signal originates solely from the sample surface, but spectra obtained for $H_2O$ are indistinguishable from those presented in Fig. 1(A). A complex spectrum with 7 distinct desorption features was reproducibly observed from several different single crystal samples, and we label the peaks α, α' β, γ, δ, ε and φ in order of ascending temperature. A plot of the integrated peak area versus exposure (Fig. 1(B)) yields a straight line, consistent with the measured sticking probability of unity at all coverages (Fig. S1). The onset of multilayer ice desorption (peak α at 155 K) occurs for a coverage close to 8.5 molecules per ($\sqrt{2}\times\sqrt{2}$)R45° unit cell (8.5 $H_2O$/u.c. = 1.28×10$^{15}$ $H_2O$/cm$^2$), which is close to the density of an ice monolayer on close-packed metal surfaces, and we thus consider everything desorbing at higher temperatures a constituent of the first water monolayer. The saturation of the φ (550 K) and ε (310 K) peaks (inset, Fig. 1(A)) occurs for coverages significantly less than 1 $H_2O$/u.c., and we assign these states to surface defects. Peaks β, γ, and δ saturate at coverages close to 8, 6, and 3 $H_2O$/u.c., respectively, which suggests that stable surface phases are completed at these coverages. α' is a small shoulder between the saturation of the β peak and the onset of multilayer desorption (peak α).

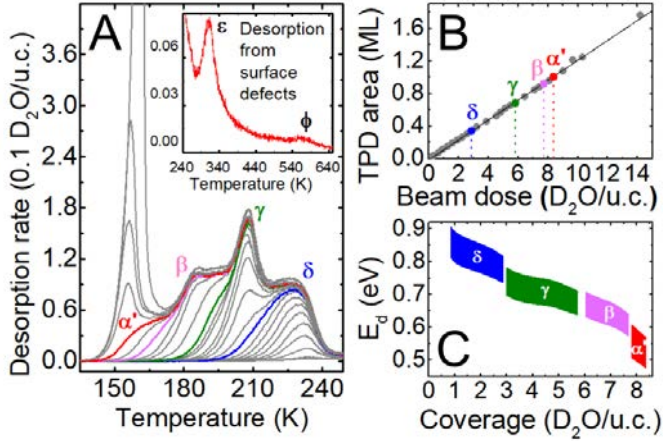

**Figure 1: Quantification of water adsorbed on Fe$_3$O$_4$(001) by TPD.** (A) Experimental TPD spectra obtained for initial D$_2$O coverages ranging from 0 to 14 molecules per Fe$_3$O$_4$(001)-($\sqrt{2}\times\sqrt{2}$)R45° unit cell (inset: higher temperature range showing desorption peaks ε and φ, which originate from surface defects). The colored curves indicate the coverages for which a particular desorption feature (labeled α', β, γ, δ) saturates. (B) Plot of the integrated TPD peak areas as a function of beam exposure. The colored data points correspond to the colored curves in panel (A). Based on these data we conclude the β, γ, and δ peaks saturate at coverages of 8, 6, and 3 molecules per ($\sqrt{2}\times\sqrt{2}$)R45° unit cell, respectively. (C) Inversion analysis of the TPD data for D$_2$O on Fe$_3$O$_4$(001) for the different peaks. The filled area marks the uncertainty range of the coverage-dependent desorption energies for each peak.

To extract information regarding the desorption energetics from the TPD data, we performed an inversion analysis [14]. Full details are contained in the supporting information. Briefly, the analysis assumes that the desorption follows first-order Arrhenius kinetics, and yields the coverage-dependent activation energy for desorption, $E_d$ (Fig. 1(C)) by direct inversion of the well-known Polanyi-Wigner equation. The uncertainty in $E_d$ is related to the uncertainty in the pre-exponential factor, $\nu$, which is unknown, but optimized during the analysis [14]. The resulting $E_d$ is equivalent to the adsorption energy ($E_{ad}$) if the adsorption is a reversible process with no activation barrier. The results, shown in Fig. 1(C), show that $E_d$ decreases with increasing coverage from a maximum of 0.85 ± 0.05 eV in the low-coverage limit to 0.52 ± 0.05 eV for the first molecules desorbing from the first monolayer. Interestingly, the corresponding values of $\nu$ ($\nu_\delta=10^{17\pm1}$ s$^{-1}$, $\nu_\gamma=10^{16\pm1}$ s$^{-1}$, $\nu_\beta=10^{16\pm1}$ s$^{-1}$, $\nu_{\alpha'}=10^{14\pm2}$ s$^{-1}$) are relatively high, which suggests that the adsorbed state is highly constrained [14]. For comparison, utilizing $\nu = 10^{13}$ s$^{-1}$ (appropriate for a 2D gas) in Fig. 1(C) would see all desorption energies lowered by approximately 0.15 eV.

To understand the origin of the complex multi-peak desorption profile we studied the adsorbed water structures with STM and nc-AFM. Figure 2(A) shows an STM image of the as-prepared Fe$_3$O$_4$(001) surface acquired at 78 K. Rows of protrusions in the [110] direction are due to the octahedrally-coordinated surface Fe$_{oct}$ atoms of a stoichiometric surface layer (see surface model in inset). The surface oxygen atoms are not imaged because there are no O-related states in the vicinity of the Fermi level. The undulating appearance of the Fe$_{oct}$ rows and associated ($\sqrt{2}\times\sqrt{2}$)R45° periodicity (white square) are linked to a subsurface rearrangement of the cation sublattice [15]. We have previously shown that the O* atoms (i.e. surface oxygen without a tetrahedrally coordinated Fe$_{tet}$ neighbor in the second layer) are active sites for adsorption. These atoms differ electronically from the others (DFT predicts a small magnetic moment [15]), and they stabilize metal adatoms to high temperatures [16]. Crucially for what follows, the O* atoms are also preferred sites for the formation of O$_{surface}$H groups [17,18]. There is always a small coverage of O*H following in-situ preparation due to the reaction of water from the residual gas with oxygen vacancies, and they cause pairs of surface $Fe_{oct}$ protrusions to be imaged slightly brighter in empty states STM images. An example is highlighted by a white arrowhead in Fig. 2(A). Recombination of the O*H species with lattice O to desorb water is responsible for the φ peak (550 K) observed in TPD [19].

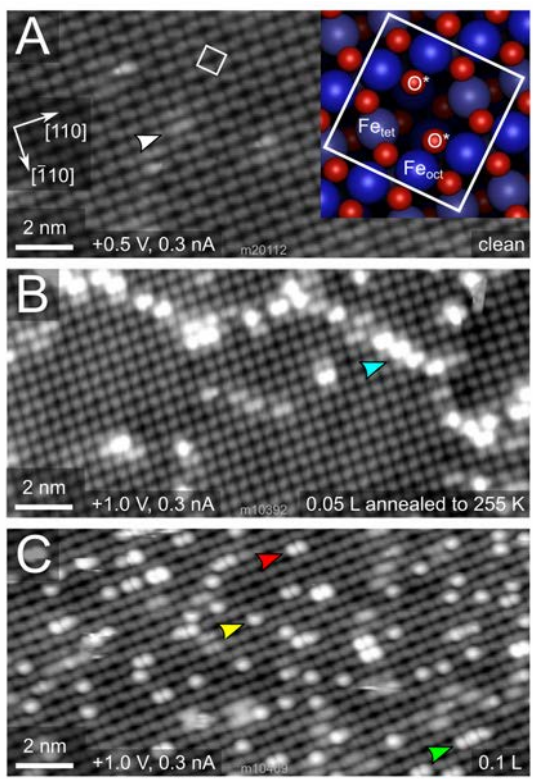

**Figure 2: Water monomers, dimers and chains on the $Fe_3O_4$(001) surface imaged by low-temperature (78 K) STM** (A) The as-prepared $Fe_3O_4$(001) surface. The (√2×√2)R45° periodicity is indicated by the white square, and the white arrow highlights an O*H group. (inset) Top view of the $Fe_3O_4$(001)-(√2×√2)R45° surface structure with the subsurface cation vacancy structure. Only the $Fe_{oct}$ atoms are imaged in STM. (B) STM image acquired after 0.05 L water adsorbed and heated to 255 K. The surface is clean, except for protrusions located at surface defects including antiphase domain boundaries in the (√2×√2)R45° reconstruction (cyan arrow). (C) STM image following adsorption of 0.1 L water at 120 K. Isolated single protrusions (yellow arrow), double protrusions (red arrow) and longer chains (green arrow) are due to water molecules adsorbed on the $Fe_{oct}$ rows.

To confirm the ε TPD peak at 310 K was defect related, we exposed the as-prepared $Fe_3O_4$(001) surface to 0.05 L water, heated to 255 K, and imaged the surface using STM. Figure 2(B) shows bright protrusions adsorbed at an antiphase domain boundary in the (√2×√2)R45° reconstruction [20], and there is also evidence for adsorption at $Fe^{2+}$ related point defects and step edges (Fig. S3). Similar behavior was recently observed for methanol on this surface [21]. In the current paper, we are primarily interested in water adsorbed at regular lattice sites.

Figure 2(C) shows an STM image of the $Fe_3O_4$(001) surface after exposure to 0.1 L (1 L = 1.33x$10^{-6}$ mbar.s) $H_2O$ at 120 K. At this temperature, far below the desorption threshold, surface mobility is low, and we observe a non-equilibrium state. The image, acquired at 78 K, exhibits isolated, bright protrusions on the $Fe_{oct}$ rows due to adsorbed water (yellow arrow). It is not straightforward to determine whether the molecules are intact or dissociated from this image, but several water dimers are observed already at this coverage (red arrow). Interestingly, dimers have two apparent heights, so there may be two types of water dimers under these conditions.

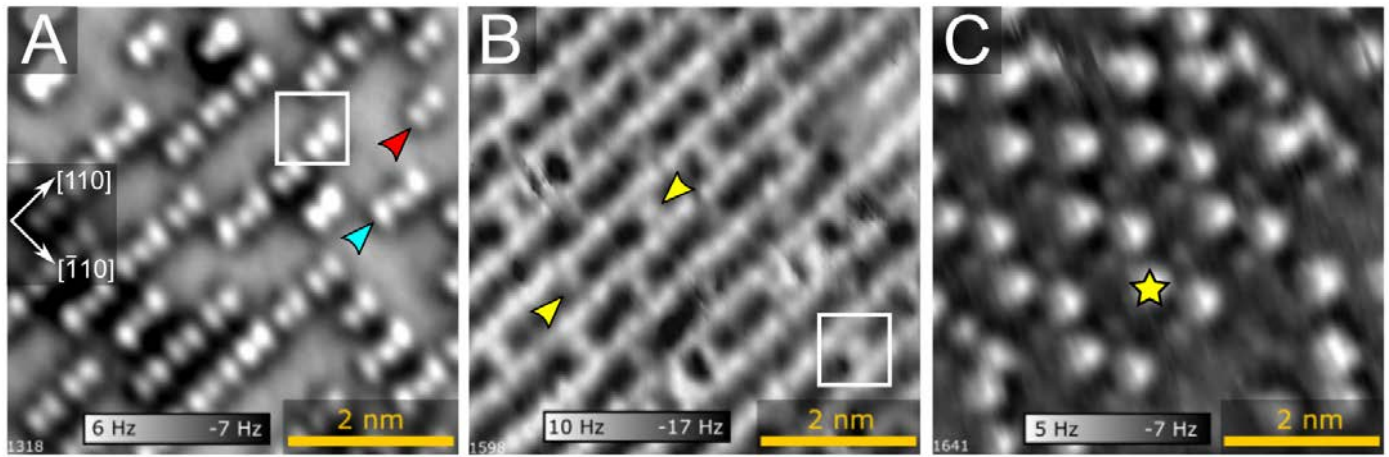

**Figure 3: Imaging water agglomerates on $Fe_3O_4$(001) with nc-AFM using a CO-functionalized tip.** Nc-AFM images obtained after exposing the as-prepared $Fe_3O_4$(001) surface to (A) 2.5 ± 0.5 $H_2O$/u.c., (B) ≈6 $H_2O$/u.c. and (C) ≈8 $H_2O$/u.c. In each case, water was dosed at 105 K, and the sample preheated to ≈155 K prior to imaging at 78 K. The coverages in panels (A), (B), and (C) correspond roughly to saturation of the δ, γ, and β peaks in TPD, respectively. Partially-dissociated water dimers and trimers on the $Fe_{oct}$ rows are indicated by red and cyan arrows in (A), respectively, and yellow arrows highlight protrusions bridging the $Fe_{oct}$ rows in panel (B). Additional water deposited on the surface appears as bright protrusions (yellow star), suggesting it protrudes significantly above the ring-like structure (C). The (√2×√2)R45° surface unit cell is shown by a white square.

Finally, there are instances of longer water chains (green arrow), but it is difficult to know how much water is involved, and these could simply be two dimers. Nevertheless, the STM data suggest that water molecules can diffuse already at 120 K, and interact attractively should they meet. STM images of higher water coverages were acquired (see Supplement), revealing limited additional information. The $Fe_{oct}$ rows are increasingly occupied by extended protrusions, but it is not possible to resolve the internal structure (Fig. S3).

To learn more about water in the sub-monolayer regime, we imaged the surface using nc-AFM. The best images were obtained in constant-height mode using a CO-functionalized tip (Fig. 3). This experimental setup was recently utilized to image water clusters on different surfaces [12,22–24], and the observed image contrast was attributed to electrostatic interaction between the CO quadrupole field and strongly polar water molecules [24]. This mechanism provides stable, molecular resolution at relatively large tip-sample distances, where the tip does not interact with the water clusters.

Figure 3(A) shows an nc-AFM image of the $Fe_3O_4$(001) surface after 2.5 ± 0.5 $H_2O$/u.c. $H_2O$ was adsorbed at 105 K. Prior to imaging, the sample was heated to 155 K, which is short of the desorption onset of the δ peak. The image, acquired at 78 K, exhibits a bi-modal distribution of double (red arrow) and triple (cyan arrow) protrusions aligned with the [110] direction, which we assign to water dimers and water trimers, respectively. The bright spots originate from repulsive electrostatic interaction between the CO tip and the O atom of the water molecule or OH group. The distance measured between neighboring protrusions within each dimer/trimer is 0.3 nm, consistent with adsorption at the surface $Fe_{oct}$ cations on the underlying surface (see structural model in Fig. 2(A)). Again, it is impossible to know from the AFM images alone whether the species within the dimer/trimer are intact or dissociated.

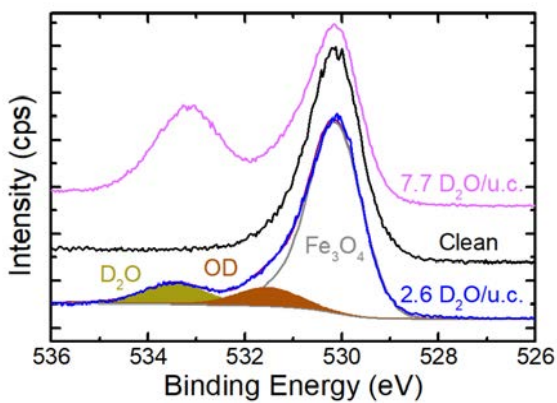

**Figure 4: O1s XPS data showing that the water agglomerates formed on $Fe_3O_4$(001) are partially dissociated.** The as-prepares surface exhibits a single peak at 530.1 eV due to the lattice oxygen atoms. The 2.6 $D_2O$/u.c. data should be compared with the surface shown in Fig. 3(A), and shows roughly equal contributions from OD and $D_2O$, consistent with one dissociated molecule per water dimer/trimer. Most of the additional water adsorbed at a coverage of 7.7 $H_2O$/u.c. is molecular. Data were measured at 95 K, with monochromatic Al Kα radiation and at a grazing exit of 80° for the emitted photoelectrons.

Figure 3(B) was acquired after the water coverage was increased to ≈6 $H_2O$/u.c., and the sample again heated to 155 K prior to imaging at 78 K. The image exhibits full rows of bright protrusions along [110]; four protrusions are observed per unit cell, consistent with adsorption on all surface $Fe_{oct}$ atoms. In addition, protrusions are observed in between the rows (yellow arrows). In most cases, the distance between these bridging protrusions along [110] is 1.19 nm, which corresponds to the periodicity of the (√2×√2)R45° reconstruction in that direction. Finally, when the coverage is increased (Fig. 3(C)), the contrast becomes dominated by new features, which protrude further from the surface than the rest of the water layer. This suggests that there are additional stable binding sites available on the 6 $H_2O$/u.c. structure, or that the layer restructures above this coverage.

To ascertain the chemical state of the water within the adlayers we performed XPS experiments. Figure 4 shows the O 1s region for the as-prepared $Fe_3O_4$(001) surface (black curve), and after 2.6 (blue curve) and 7.7 (magenta curve) $D_2O$/u.c. was adsorbed and the sample was heated to 175 K with a 1 K/s ramp. The as-prepared surface exhibits a single, slightly asymmetric peak at 530.1 eV due to the lattice oxygen [25]. Exposure to water creates a clear peak at 533.4 eV due to $D_2O$, which shifts slightly to lower binding energy with increasing coverage. Fitting the 2.6 $D_2O$/u.c. data with Voigt functions, we find that (at least) one additional peak at 531.5 eV is required to accurately model the data. This peak position is close to that observed previously for $O_{surface}H$ groups (531.3 eV) [26]. Of course, the XPS binding energy of $O_{water}H$ groups could be slightly different, particularly when it is part of an agglomerate, but calculated core level shifts [27] for the $O_{surface}H$ and the $O_{water}H$ of the linear water trimer (see Fig. 5) found a difference of 0.1 eV. Since $D_2O$ dissociation yields two OD groups, the similar peak areas at 533.4 eV and 531.5 eV suggests that approximately half of the $D_2O$ is dissociated. At the higher coverage, the area of the $D_2O$ peak increases significantly, and shifts to lower binding energy. The peak area in the OD region remains constant with respect to the substrate peak (fit not shown), which suggests that the additional water adsorbs molecularly.

To understand the formation of different water structures we now turn to our computational results. As explained below, we employed a systematic approach to determine the lowest energy configuration of water molecules in the coverage regime 0-8 $H_2O$/u.c. It is important to note that this is not an automated genetic algorithm, but rather proceeds by identifying factors that certain trial structures more stable than others at each coverage, and using this information to build subsequent generations. A complete account of the theoretical approach, and discussion of all structures computed will be published separately. Selected results relevant to the discussion here are shown in Fig. 5. Before continuing, it is important to note that our calculations utilize the GGA+U approach ($U_{eff}$=3.61 eV) [28,29] with the optPBE-DF exchange-correlation (Xc)-functional [30–32] which is modified to include long-range vdW interactions, and the so-called subsurface cation vacancy (SCV) model of $Fe_3O_4$(001)-($\sqrt{2}\times\sqrt{2}$)R45° [15]. Thus, our setup differs markedly from the prior work of Mulakuluri et al. [5,33], who utilized a standard GGA+U functional and a bulk-truncated surface model, and only calculated coverages of 1, 2 and 4 $H_2O$/u.c..

Interestingly, we find that an isolated water molecule prefers to adsorb molecularly on the $Fe_3O_4$(001) surface (Fig. 5(A), $E_{ad}$=-0.64 eV). The optimum configuration has the O atom close to atop a substrate $Fe_{oct}$ cation, with the molecule in the plane of the surface and oriented such that the H atoms interact with nearby surface O atoms via very weak H bonds, 2–2.2 Å. This configuration is 0.05 eV more stable than a dissociated molecule ($E_{ad}$=-0.59 eV), where the OH group adsorbs upright atop a $Fe_{oct}$ cation, with the proton deposited at the neighboring O* forming a surface hydroxyl (O*H).

The most stable configuration of water on the $Fe_3O_4$(001) surface occurs at a coverage of 2 $H_2O$/u.c. with the formation of a partially dissociated water dimer ($E_{ad}$=-0.92 per molecule). This species comprises one terminal OH and one $H_2O$, bound to neighboring surface $Fe_{oct}$ atoms along the row, connected by an inter-molecular H-bond (1.41 Å). The $H^+$ atom liberated by the dissociation forms an O*H group. Further details of the adsorption geometry are included in the discussion section, where we explain the cooperative origin of this species' stability.

The DFT-based search at 3 $H_2O$/u.c. yields two partially dissociated water trimers degenerate in energy (see Fig. 5(A), labeled $E_3$ and $E_{3\,ISO}$). Both species are based on the partially-dissociated water dimer described above, but differ in the location of a third molecule. In the linear $H_2O$-OH-$H_2O$ trimer, the third molecule binds on the surface $Fe_{oct}$ row, and donates an H-bond into the OH. In the alternative non-linear isomer trimer, the third water molecule binds by H-bonds only. It receives an H-bond from the surface O*H, and donates an H-bond to the nearby, unoccupied O* atom. Electrostatic repulsion renders the adsorption of a proton at both O* sites energetically unfavorable at low coverage, and thus dissociation is limited to one molecule per ($\sqrt{2}\times\sqrt{2}$)R45° unit cell.

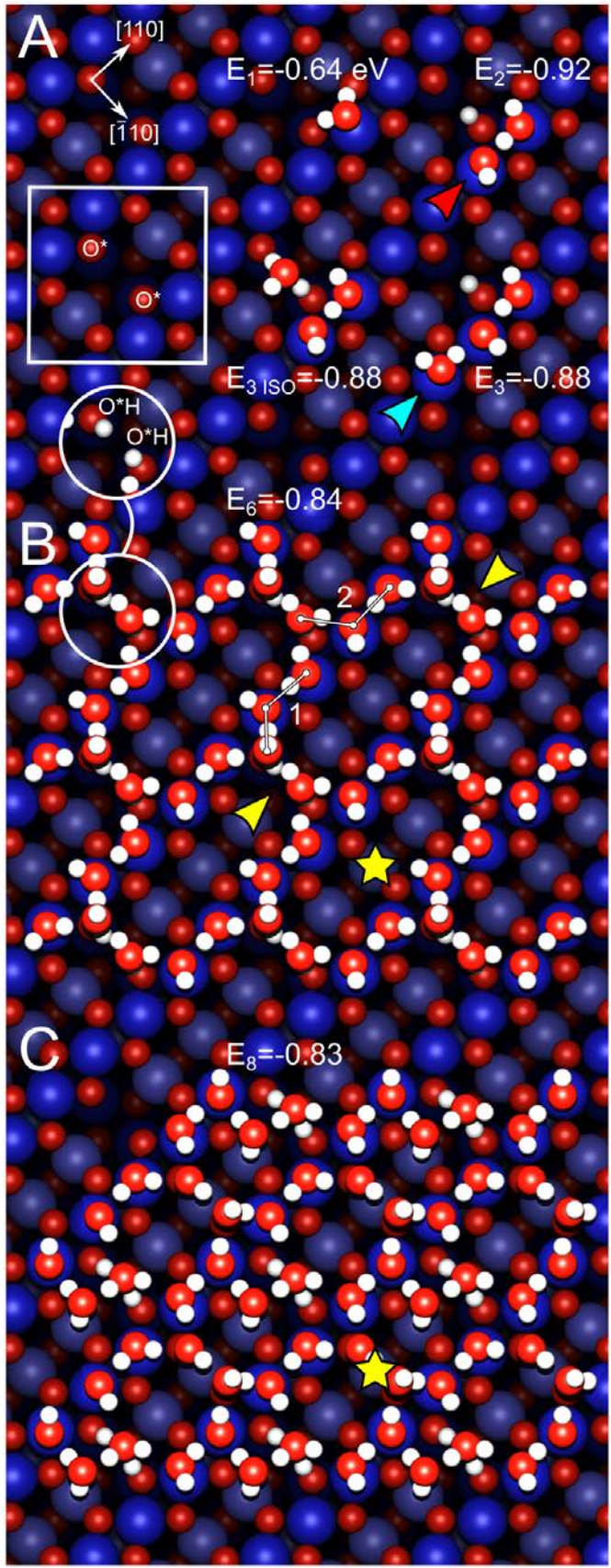

Figure 5: Top view of the minimum-energy structures determined by DFT for water coverages of 1, 2, 3, 6 and 8 $H_2O$/u.c.. (A) An isolated molecule adsorbs intact, but partially-dissociated water dimers and trimers are energetically preferred. Two partially-dissociated trimer structures are calculated to be energetically degenerate. Fe atoms are blue, and O are red. (B) DFT-based model at 6 $H_2O$/u.c. showing a ring-like structure based on full occupation of the $Fe_{oct}$ rows with OH or $H_2O$, and water molecules bridging the O* sites. These bridging molecules are adsorbed partly through H-bonds to surface O*H groups. The O*H groups beneath the adsorbed molecules are shown in the rightmost white circle. Alternatively, the structure can be viewed as based on a pair of $H_2O$-OH-$H_2O$ timers (labeled 1 and 2). (C) DFT-based model at 8 $H_2O$/u.c. showing a complex structure utilizing dangling bonds in the 6 $H_2O$/u.c. structure to form a second bridge in the region of the yellow star. All adsorption energies are given in eV. The ($\sqrt{2}\times\sqrt{2}$)R45° unit cell and both O* are highlighted.

The lowest-energy structure determined by our DFT search at 6 $H_2O$/u.c. exhibits a ring-link appearance, in agreement with the nc-AFM image shown in Fig. 3(B). This is the first coverage at which all adsorbed molecules are involved in a H-bonded network that covers the surface. All four $Fe_{oct}$ sites in each ($\sqrt{2}\times\sqrt{2}$)R45° unit cell are occupied by either $H_2O$ or OH, and the rows are bridged by two further water molecules attached solely through H-bonds. In general, the structure is characterized by $H_2O$-OH-$H_2O$ trimers, and facilitates near-ideal bonding angles of 122-124° for intact water molecules. Interestingly, the repulsive behavior of the two O*H species observed at lower coverage is mitigated through the additional H-bonding with the bridge molecules. The structure at 8 $H_2O$/u.c. is shown in Fig. 5(C). It is rather complex, but essentially water utilizes the remaining dangling H-bonds in the 6 $H_2O$/u.c. structure to form a second bridge of molecules near the center of the previously ring-like feature (in the vicinity of the yellow star in Fig. 5(B)). However, additional reorganisation occurs to optimise the H-bonding, including a modification of the original bridge structure formed at 6 $H_2O$/u.c.. Since the coverage at 8 $H_2O$/u.c. is already close to that of a close-packed ice layer, it is straightforward to understand why further water adsorption results in the adsorption of multilayer ice. All structures shown in Fig. 5 can be downloaded as part of the SI.

**Discussion**

Based on the experimental and theoretical evidence presented above, we conclude that partially-dissociated water dimers are the most stable species on the $Fe_3O_4$(001) surface, closely followed by structurally related, partially-dissociated water trimers. Our nc-AFM images clearly show the adsorbed dimers and trimers, and XPS spectra reveal them to be partially dissociated. Moreover, the theoretically determined adsorption energies agree remarkably well with the $E_d$ values obtained from an inversion analysis of the δ-peak, and the highly constrained adsorption geometry predicted by DFT is consistent with the high pre-exponential factor ($\nu$ = $10^{17\pm1}s^{-1}$). For higher coverages, the inversion analysis reveals the $E_d$ necessary to desorb the most weakly bound molecule(s), and thus should not be compared to the average adsorption energies calculated by DFT.

Clearly, the (√2×√2)R45° reconstruction plays a crucial role in the adsorption behavior. At low coverages, the partially-dissociated water dimers and trimers order with (√2×√2)R45° symmetry, while at high coverages the structure of the H-bonded network also belies the periodicity of the underlying substrate. Ultimately, this stems from the strong preference to form surface O*H groups, which limits the density of dimers/trimers to one per unit cell. Later, the O*H groups provide a hydrogen bond to bridge the $Fe_{oct}$ rows and complete the H-bonded network.

Despite the importance of the O* sites, the primary contribution to the adsorption energy at low coverage arises from the $Fe^{3+}$-$O_{water}$ bond. An isolated water molecule binds strongly atop an $Fe_{oct}$ row atom (-0.64 eV), and prefers this state to dissociation by 0.05 eV. This result differs from the prior calculations of Mulakuluri et al. [5,33], and after extensive testing, we have found that the discrepancy originates in the structural model used, and not the functional applied. As suggested by Mulakuluri et al. [33], the $Fe^{2+}$ cations in the subsurface layers of a bulk-truncated structure interact with the adsorbates and promote dissociation. The SCV reconstruction contains only $Fe^{3+}$ cations in the outermost 4 layers, and molecular adsorption is preferred.

Given the lack of dissociation in the monomer case, it is somewhat surprising that partially-dissociated water dimers form on the $Fe_3O_4$(001) surface. Recently, Freund's group [8] proposed that partial-dissociation requires two molecules to meet on the $Fe_3O_4$(111) surface, but later revised their IRAS analysis in favor of the "traditional picture" where dissociation occurs first in isolation on an under-coordinated anion-cation pair [34]. Our STM images show that dimerization occurs already at very low coverages on $Fe_3O_4$(001), and there is no evidence for monomer dissociation in the form of additional isolated O*H groups. It is, however, difficult to know if the dimers are molecular or partially dissociated from STM alone. This ambiguity does not exist for higher coverages: nc-AFM images of 2 $H_2O$/u.c. (Fig. 3(A)) show dimerization occurs already at 155 K, and analysis of XPS spectra suggests that roughly one molecule per agglomerate is dissociated.

What then, drives the partial dissociation of water dimers in the water/$Fe_3O_4$(001) system? To answer this question we first analyze the DFT results for a molecular water dimer (Fig. 6(A)). Somewhat surprisingly, the energy gain of molecular dimerization is small; an isolated molecule has a binding energy of -0.64 eV, while the average binding energy in the molecular dimer is just -0.66 eV per molecule. This difference is significantly less than the binding energy of an H-bond in a gas-phase water dimer (-0.10 eV per molecule). Since the H-bond length in the present system (1.89 Å) is significantly shorter than that of a gas-phase water dimer (2.0 Å), some energy must be lost in the final structure. In this regard, we consider the $Fe_{oct}$-$O_{water}$ bond lengths. The water that donates an H-bond has an $Fe_{oct}$-O bond of 2.20 Å, comparable to the isolated water monomer (2.22 Å), but the acceptor molecule has an $Fe_{oct}$-O bond length of 2.34 Å. This suggests that receiving an H-bond weakens the interaction of a water molecule with the substrate, consistent with the idea that forming $Fe_{oct}$-O bonds and receiving H-bonds both involve the lone pair (O 2p) orbitals [35], and that the competition leads to the marginal total-energy gain.

The situation is very different for a partially-dissociated water dimer (Fig. 6(B)). Here, the average adsorption energy per molecule is -0.92 eV, so the small energetic cost of dissociating one molecule (0.05 eV for an isolated monomer) is easily compensated. The intermolecular H-bond is significantly shorter (1.41 Å) in the partially-dissociated water dimer, as expected on electrostatic grounds (the OH species is negatively charged). Moreover, the H-bond donating water molecule has a significantly shorter $Fe_{oct}$-$O_{water}$ bond (2.06 eV) than in the molecular water dimer, suggesting a stronger interaction with the substrate. This phenomenon has been observed in gas-phase water clusters [36], and on metal surfaces [35,37], and is known as cooperativity [38]. Essentially, water molecules seek a balance in their H-bonding interactions. If a molecule donates a strong H-bond, it accepts stronger H-bonds. Since H-bond acceptance utilizes the same O orbital as the $Fe_{oct}$-$O_{water}$ bond, the balance can be achieved through an enhanced interaction with the substrate. Thus, the formation of the negatively charged OH group induces water to donate a strong H-bond, which in turn induces a stronger water-surface interaction. The energy gain is so substantial that the system can accommodate a weakened terminal-OH $Fe_{oct}$-O bond, which is 0.12 Å longer than for an isolated terminal OH (not shown).

Next we turn our attention to the partially-dissociated water trimer (Fig. 6(C)). The linear trimer is a natural consequence of the arguments outlined above, as there is an under-coordinated $Fe_{oct}$ atom available to which a water molecule can bind and simultaneously donate an H-bond into the OH group of the partially-dissociated dimer. Of course, the Fe-Fe separation along the $Fe_{oct}$ rows is considerably larger (3 Å) than the sum of OH and H-bond lengths, so the partially dissociated water trimer forms with the OH group directly atop an $Fe_{oct}$ atom, with both water molecules leaning in toward the OH from their favored position atop an $Fe_{oct}$ (see Fig. 6). The H-bonds are slightly longer, as is the OH $Fe_{oct}$-O bond. This results in the slightly lower adsorption energy of -0.88 eV per molecule compared to the OH-$H_2O$ water dimer.

To this point in the discussion, the DFT search predicts what is observed in experiment, and allows us to understand why the partially-dissociated water agglomerates form, and are so strongly bound. However, DFT finds a non-linear partially dissociated water trimer ($E_{3\,ISO}$ in Fig. 5(B)) with a comparable adsorption energy to the partial dissociated linear trimer, which is not observed in the experiments. The fact that we can resolve water bridging the $Fe_{oct}$ rows upon the formation of the H-bonded network with nc-AFM in Fig. 3(B) gives us confidence we could detect the non-linear trimer if it were present. This indicates that the energy balance involved in adsorbed water agglomerates is not perfectly handled by DFT. To investigate further, we compared several alternative functionals, with and without vdW corrections, and obtained similar trends, albeit with a variation of +/- 0.1 eV in the absolute energies. We conclude that DFT is insufficiently accurate to predict the relative stabilities of water adlayers, and that calculating reliable structures requires guidance from experiment or the adoption of superior approaches such as hybrid functionals or the random phase approximation [39,40].

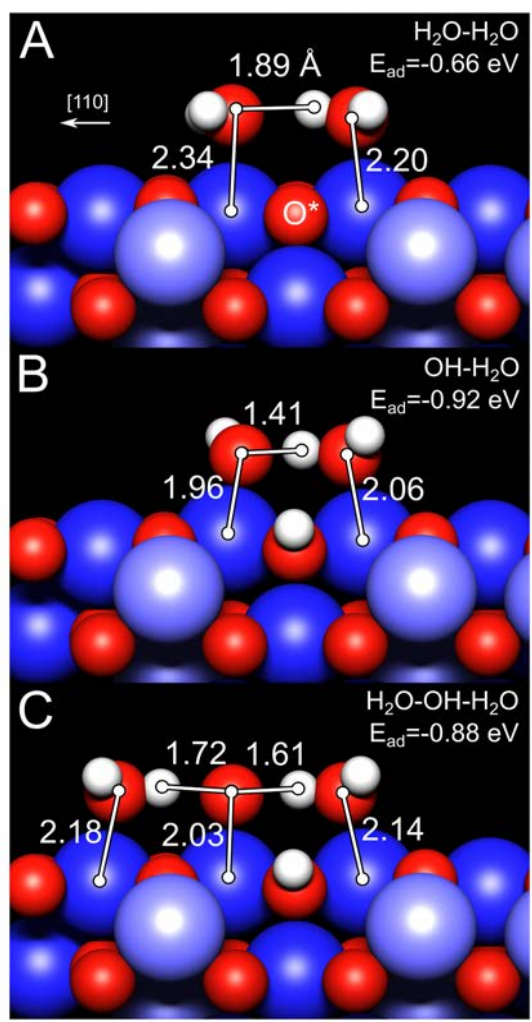

**Figure 6: The geometry of partially-dissociated water dimers and trimers reveals a cooperative binding effect.** (A) A molecular water dimer exhibits a relatively long inter-molecular H-bond, and the H-bond acceptor has a weakened interaction with the surface compared to an isolated molecule. (B) The partially-dissociated water dimer exhibits a strong inter-molecular H-bond, and the H-bond-donating water molecule binds more strongly to the substrate. (C) In the partially-dissociated water trimer, as second water molecule donates an H-bond to the OH group, further weakening its bond to the substrate. All bond lengths are given in Å and energies in eV.

Once the partially-dissociated water trimer (Fig. 6(C)) forms, it is not possible to H-bond additional water along the $Fe_{oct}$ row. The TPD data suggests that the next stable structure occurs at a coverage of 6 $H_2O$/u.c., where the nc-AFM images (Fig. 3(B)) clearly resolve a ring-like structure with additional protrusions bridging the molecules adsorbed at the $Fe_{oct}$ rows. This is in line with the minimum-energy structure determined at 6 $H_2O$/u.c., which is the first configuration to establish an H-bonded network extending over the whole surface. The stability of the proposed structure stems from the presence of $H_2O$-OH-$H_2O$ trimers, which utilize $H_2O$ molecules stabilized at the O* bridge sites. We note however, that the experimental data does not allow to unambiguously confirm the fine details of the structure, and that the XPS fitting at 7.7 $D_2O$/u.c. suggests that less water is dissociated. The same is true at a coverage of 8 $H_2O$/u.c., the next coverage where the TPD data indicates that a stable structure exists. The nc-AFM images show that new, ordered protrusions emerge when additional water is added to the ring-like structure (Fig. 3(C)), and DFT predicts that this water binds via the remaining dangling H-bonds present in the 6 $H_2O$/u.c. structure. That these molecules bind solely by H-bonds to other water explains why the desorption temperature is so close to that of multilayer ice. The reason for the observed strong AFM contrast is not yet known, and we cannot discount the possibility of additional rearrangement at this coverage. What is clear, is that the β and α' peaks observed in TPD are due to water squeezed into the 6 $H_2O$/u.c. structure, leading to a re-optimization of the available H-bonds.

As mentioned above, partially dissociated water dimers have recently been reported to be the most stable species on $RuO_2$(110) [7] and $Fe_3O_4$(111) [8,9], and appear to be common on metal oxide surfaces. Based on the analysis presented here, we expect these species to form whenever there are under-coordinated surface cations sufficiently close together that a $H_2O$-OH bond can be established, provided there are under-coordinated O atoms available to form a stable surface hydroxyl. Here, the SCV reconstruction of $Fe_3O_4$(001) limits the availability of the latter sites, which is why the water-dimer coverage saturates at one per unit cell. On $RuO_2$(110) [7], for example, where the under-coordinated surface O atoms are homogeneous and plentiful, a complete coverage of $H_2O$-OH dimers is achieved.

Before concluding, it is worth to consider whether the low temperature/low pressure phenomena observed here bear any resemblance to the adsorption/desorption of water on $Fe_3O_4$(001) under more realistic conditions. At first glance, the adsorption threshold of $10^{-2}$ mbar observed by Kendelewicz et al. [26] at room temperature in ambient-pressure XPS studies suggests a significant pressure gap. However, this threshold is entirely consistent with our assertion that isolated molecules are weakly bound, and that a strongly bound partially-dissociated water dimer species forms when two molecules meet on the surface. Given the binding energy of the water monomer (-0.64 eV) determined here, the $10^{-2}$ mbar threshold corresponds to an instantaneous coverage of 0.2 $H_2O$/u.c. at 273 K. This is sufficient to expect that two monomers can meet before desorbing. The as-formed dimer is more strongly bound, so a stable coverage will develop rapidly. Alternatively, the $10^{-2}$ mbar threshold corresponds to a chemical potential of −0.78 eV, which agrees very well with the adsorption energy of the partially dissociated water dimer determined by the inversion analysis (-0.82 eV). As such, the surface science approach utilized here appears directly applicable to understand the adsorption/desorption of water at pressures relevant to catalysis. To our knowledge, the reactivity of partially-dissociated water dimers have not been studied directly, and it will be fascinating to see if these species play an active role in geochemical or corrosion processes, or in catalysis where metal oxides are frequently used as a catalyst or as a support for metal nanoparticles. In particular, it will be interesting to learn whether partially dissociated species play a role in the water-gas shift reaction, because industry currently utilizes an $Fe_3O_4$ based catalyst [41–44], and partially dissociated species have now been directly identified on both major facets.

In summary, the formation of partially-dissociated water agglomerates on $Fe_3O_4$(001) is driven by the formation of strong intermolecular H-bonds and facilitated by the close proximity of under-coordinated cations. The presence of the SCV reconstruction ensures that partially-dissociated water dimers and trimers remain isolated because there is only one O site that can accommodate a proton per unit cell. The partially-dissociated agglomerates act as an anchor to build a ring-like H-bonded network as the coverage is increased, and the water layer completes by saturating dangling H-bonds within this stable structure. A similar evolution can be expected wherever a surface presents well-spaced active sites for dissociation.

ACKNOWLEDGEMENTS: The authors gratefully acknowledge funding through projects from the Austrian Science Fund FWF (START-Prize Y 847-N20 (MM, JH, RB & GSP); Special Research Project 'Functional Surfaces and Interfaces', FOXSI F4505-N16 and F4507-N16 (MS & UD)), the European Research Council (UD: ERC-2011-ADG_20110209 Advanced Grant 'OxideSurfaces'), and the Doctoral College TU-D (ZJ) and Solids4fun (W1243: RB). The computational results presented have been achieved using the Vienna Scientific Cluster (VSC).

**Methods**

The TPD and XPS experiments were performed in a vacuum system optimized for the study the surface chemistry of metal oxide single crystals. The system has been described in detail elsewhere [13]. Briefly, the single crystal $Fe_3O_4$ sample (6x6x1 mm, SurfaceNet GmbH) is mounted on a Ta backplate in thermal contact with a L-He flow cryostat. The sample can reach a base temperature of ≈30 K, and can be heated to 1200 K by direct current heating of the sample plate. Temperatures are measured by a K-type thermocouple spot welded to the sample plate, and calibrated by the multilayer desorption of simple gases [45]. $D_2O$ was adsorbed directly onto a 3.5 mm diameter spot in the centre of the sample surface using an effusive molecular beam source. The beam has a close to top-hat profile and has a precisely calibrated flux (9.2 ± 0.5 × $10^{12}$ $D_2O$ molecules/$cm^2$.s) at the sample. Coupled with sticking-probability measurements, this allows accurate prediction of the absolute water coverage [13,46]. This is particularly straightforward to achieve here, because the sticking probability for water is unity at all coverages at 100 K (see Fig. S1 and Fig. 1(B)). For TPD experiments the sample is exposed to water at 100 K, and then heated with a linear ramp of 1 K/s. XPS utilizes a SPECS Phoibos 150 analyser with a monochomatized FOCUS 500 Al Kα X-ray source. STM and nc-AFM measurements were performed in a separate vacuum system using an Omicron LT-STM equipped with a QPlus sensor. Here, water exposures were performed using a high-precision leak valve. The water coverage is defined in $H_2O$ molecules per $(\sqrt{2}\times\sqrt{2})R45°$ unit cell ($H_2O$/u.c.), where 1 $H_2O$/u.c. is a coverage of $1.42\times10^{14}$ $cm^{-2}$.

The Vienna *ab initio* Simulation Package (VASP) [47,48] was used for all density functional theory (DFT) calculations. The Projector Augmented Wave (PAW) method describes the electron and ion interactions, with the plane wave basis set cut-off energy set to 550 eV. A Γ-centered k-mesh of 5×5×5 was used for bulk calculations, adjusted to 5×5×1 for (001) surface calculations. Convergence is achieved when the electronic energy step of $10^{-6}$ eV is obtained, and forces acting on ions become smaller than 0.02 eV/Å. The calculations are based on the "subsurface cation vacancy" (SCV) reconstructed model of the $Fe_3O_4$(001) surface [12]. Adsorption energies $E_{ad}$ are corrected for the zero-point energy (ZPE) (details in S.I.) and are quoted as an average (per molecule, if > 1 $H_2O$/u.c.), unless otherwise mentioned. The optPBE-DF [30–32] functional (details in S.I.) accounts for the van der Waals corrections, and ultimately delivers results that correlate well with the experiments. The same functional was recently used to simulate water adsorbed on NaCl(001) and MgO(001) surfaces [49] and water clusters [50].

The optimum configuration for each water coverage was determined via a systematic search inspired by genetic algorithms. For each coverage, the results of at least 10 trial calculations were analyzed to identify factors leading to a low total energy. These insights, together with those found at other coverages, were used to build a next generation of trial structures. This process eventually leads to the energetically lowest configuration the system can reach for the given coverage of water. In the end, over 500 configurations have been investigated.

# Water Agglomerates on Fe$_3$O$_4$(001)

Matthias Meier,[1,2] Jan Hulva,[1] Zdeněk Jakub,[1] Jiří Pavelec,[1] Martin Setvin,[1] Roland Bliem[1], Michael Schmid,[1] Ulrike Diebold,[1] Cesare Franchini,[2] and Gareth S. Parkinson[1]

[1]Institute of Applied Physics, Technische Universität Wien, Vienna, Austria

[2]University of Vienna, Faculty of Physics and Center for Computational Materials Science, Vienna, Austria

## Supporting information

### D$_2$O dosing

For both, TPD and XPS experiments, D$_2$O was dosed by an effusive molecular beam with its intensity calibrated [1]. During the dosing, D$_2$O molecules with a thermal energy of a room-temperature molecular beam were impinging at the surface with the intensity of 0.065 molecules per surface unit cell per second. Figure S1 shows the D$_2$O signal (m/e = 20) measured by a mass spectrometer positioned line of sight to the sample during dosing at 100 K (black curve) and 680 K (gray curve). The sticking coefficient of thermal kinetic energy D$_2$O molecules at 100 K is typically close to unity [2]. Therefore, all dosed molecules remain at the sample and the signal measured by mass spectrometer does not change after the molecular-beam shutter is opened (black line). Prolonged exposure of the sample to the D$_2$O beam leads to growth of water multilayers since the ice sublimation rate at 100 K is negligible. Taking advantage of the sticking coefficient being close to unity, we can directly calculate the absolute coverage for each TPD curve by multiplying the beam intensity by the dosing time.

When D$_2$O is dosed at 680 K, the mass spectrometer signal increases immediately after opening of the shutter and stays constant. This is a consequence of all available adsorption states desorbing below 650 K (inset of Fig. 1 in the main text) so molecules dosed at this temperature are scattered off the sample and can be detected by the mass spectrometer (gray line in Fig. S1).

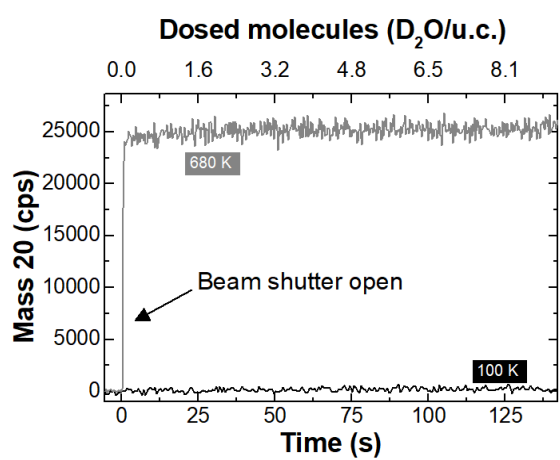

**Figure S1:** D$_2$O signal measured while dosing by molecular beam for a sample temperature of 100 K (black curve) and 680 K (gray curve), molecular beam intensity = 0.065 D$_2$O/u.c. per second.

### Inversion analysis

Details of the inversion analysis presented in Fig. 1(C) in the main text are shown in Fig. S2. TPD curves of the saturated peaks used to obtain coverage dependent desorption energies for a range of pre-exponential factors $v$ are shown in Fig. S2(A) as thick solid lines. The high-temperature parts of individual peaks belonging to the next desorption feature were cut off to perform the analysis separately for individual peaks. Desorption curves for lower initial coverages were simulated using the obtained coverage dependent desorption energies. Simulated curves were then compared with the experimental data to find the value of $v$ giving the best agreement. Examples of corresponding experimental and the best matching simulated curves are shown in Fig. S2(A) as thin solid and black dashed lines, respectively.

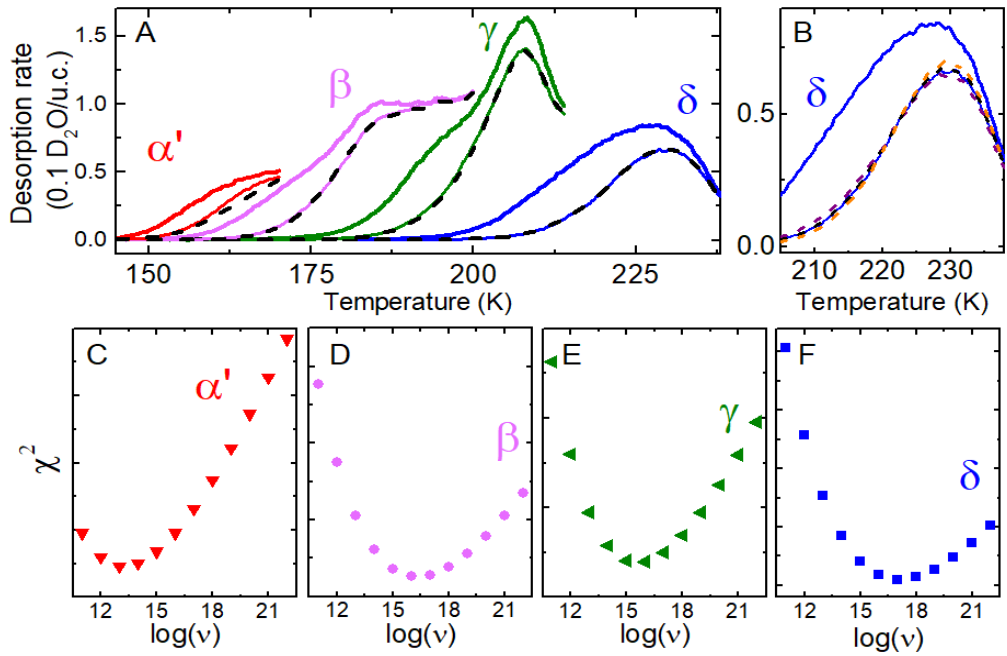

**Figure S2:** Inversion analysis of the TPD data. **(A)** Desorption curves corresponding to saturated desorption peaks (thick solid lines) were used for inversion to simulate lower coverage curves (thin solid lines, only one low-coverage curve for each peak is shown here). The black dashed lines represent the simulated curves for the best-matching pre-exponential factors. **(B)** Comparison of the experimental data (thin blue line) with the simulated data (dashed lines) from the inversion analysis for a pre-exponential factor $v=10^{15}$ s$^{-1}$ (purple), $v=10^{17}$ s$^{-1}$ (black, best fit) and $v=10^{19}$ s$^{-1}$ (orange). **(C-F)** Dependence of the total error $\chi^2$ between simulated and experimental curves on pre-exponential factor $v$. Y-axis scales between individual figures were set to display the trends of $\chi^2$.

Figure S2(B) shows the peak $\delta$ in Fig. S2(A) in detail with two additional simulated curves for $v=10^{15}$ s$^{-1}$ (dashed purple), and $v=10^{19}$ s$^{-1}$ (dashed orange). We see that these two curves differ from the experimental curve (thin solid blue) more than the best matching curve for $v=10^{17}$ s$^{-1}$ (dashed black).

To find the value of $v$ giving the best agreement we calculated the error $\chi^2$ for each peak defined as a square of the difference between experimental and simulated curve summed for all curves belonging to the given peak (except the saturated curve used for inversion) [3]. The results in Fig. S2(C-F) show the dependence of $\chi^2$ as a function of $v$, indicating an optimal value of the pre-exponential factor.

We note that the results of the analysis are less reliable for the α' peak because only one low coverage curve was available for the analysis.

The results are insensitive to which curve is chosen for the analysis. In other words, we obtain similar results if we choose a lower coverage curve instead of the curve for the saturated peak.

The temperature range for the analysis is restricted as displayed in Fig. S2(A). Trailing edges of individual peaks are 'cut-off' so as not to include these data in the error analysis. This was done to prevent the influence of the high temperature part of the curve belonging to a different desorption feature on the error analysis. Nevertheless, Fig. S2(B) shows that the biggest error caused by the variation of $v$ is located at the leading edge of the curve and on the high-temperature side of the peak apex. In addition, as the trailing edges on all TPD curves overlap at the trailing edges (see in Fig. S2(A)) the error at the high temperature side is not significant. This allow us to justify the above-described approach of the inversion analysis of the TPD spectra containing multiple peaks.

**Scanning Tunneling Microscopy**

Figure 2(A,B) (in the main text) shows a clean surface with some of the typical surface defects (anti-phase domain boundaries - APDB, surface hydroxyls - $O_{surface}H$). When water is dosed at low temperature and the sample is annealed to 255 K (above the desorption temperature of the 1$^{st}$ monolayer), water remains adsorbed at the surface defects (inset of the Fig. 1(A) in the main text). Water molecules adsorbed on surface defects are shown in Fig. 2(B) (in the main text) as intense bright protrusions. The small desorption peak φ at 550 K is straightforward to explain on the basis of our previous room-temperature study of water adsorption on $Fe_3O_4$(001) [4]. Water molecules react with a small number of surface oxygen vacancies created during sample preparation, and the O atom repairs the vacancy. The hydrogen atoms create two $O_{surface}H$ groups, and these species recombine to desorb water above 500 K leaving behind oxygen vacancies. Similar behavior is well known on reduced $TiO_2$(110) surfaces above 490 K [5]. The ε peak at 310 K is also related to desorption from surface defects such as step edges and antiphase domain boundaries (APDB) [6,7]. Bright features corresponding to $H_2O$ at the defect sites responsible for the ε peak (as seen in Fig. 2(B)) are not observed in room-temperature STM images of $Fe_3O_4$(001) because the desorption rate is extremely fast compared to the speed of an STM measurement. Since the coverage is very low, it is difficult to positively identify the presence of OH groups by XPS due to strong overlap with the O 1s peak from the $Fe_3O_4$ substrate, but the lack of molecular water in the spectrum leads us to believe that water most likely adsorbs dissociatively. This is in line with recent experiments of methanol adsorption, where dissociative adsorption (deprotonation) occurred preferentially at step edges, APDB and subsurface Fe defects, and the increased activity of these sites was linked to the presence of $Fe^{2+}$ cations [7].

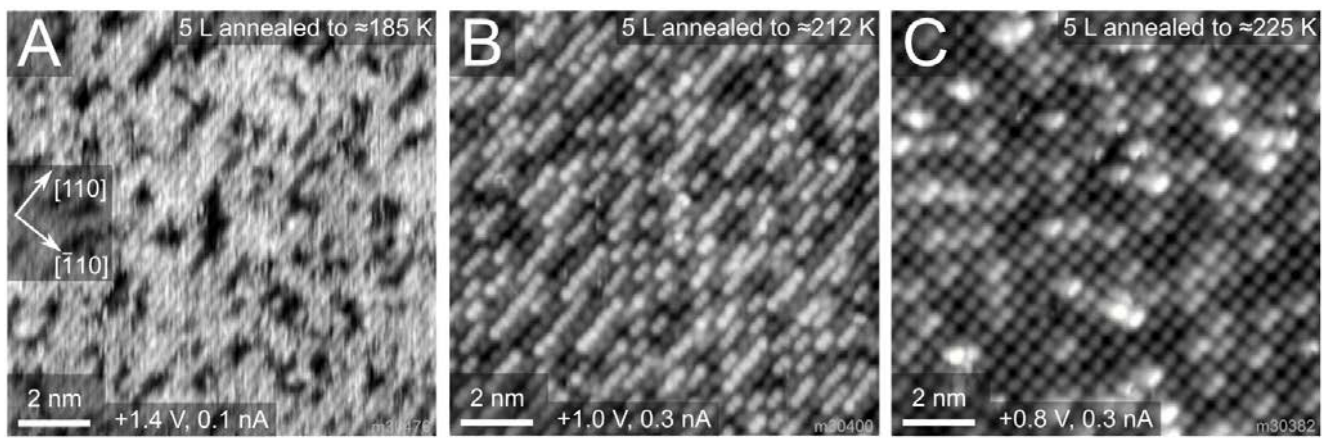

**Figure S3: :** STM image of **(A)** 5 L $H_2O$ annealed to 185 K, **(B)** 5 L $H_2O$ annealed to 212 K, **(C)** 5 L $H_2O$ annealed to 225 K. All images were acquired at 78 K.

Figures S3(A-C) show STM images (acquired at 78 K) of a surface which was exposed to 5 L (1 L = 1.33x10$^{-6}$ mbar.s) of water at 120 K and then annealed to higher temperatures in a step-wise manner to desorb part of the molecules. After annealing to 185 K for 5 minutes (Fig. S3(A)) we see fuzzy features in the direction of the surface Fe$_{oct}$ rows. After annealing to ≈212 K (Fig. S3(B)) we can clearly see chain structures in the direction of the iron rows. Their coverage strongly decreases after further annealing to ≈225 K (Fig. S3(C)) and isolated double-lobed protrusions can be seen instead. At all measured coverages the resolution of the STM does not provide any information about the inner structure of the imaged species.

**X-ray Photoelectron Spectroscopy**

X-ray Photoelectron Spectroscopy (XPS) was done using a monochromatized Al Kα radiation source. To increase the surface sensitivity, spectra were acquired at a grazing exit angle. The Fe 2p peak is not affected by water adsorption (Fig. S4). A shoulder characteristic for $Fe^{+2}$, which would indicate a change of the oxidation state of the surface Fe by a charge transfer from the proton to the surface iron atom [8], is not visible. The clean surface exhibits a slightly asymmetric O 1s peak at 530.1 eV due to the lattice oxygen [9]. Figure S4 also shows that $D_2O$ adsorbs mostly in molecular form at 40 K but partly dissociates 175 K. To rule out the influence of X-ray irradiation on the photoelectron spectra, we compared the first and the last scan of the O 1s transition. We found these to be identical, suggesting that water was not dissociated by X-rays over the time scale of the experiment.

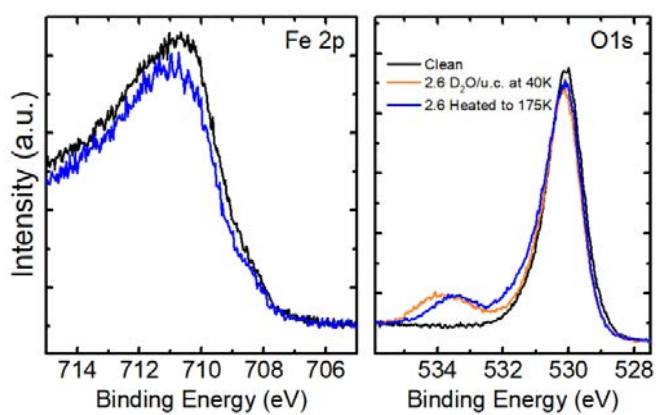

**Figure S4:** Fe 2p and O 1s of the clean surface (black), 2.6 $D_2O$/u.c. deposited at 40 K (orange, only for O 1s) and 2.6 $D_2O$/u.c. heated to 175 K (blue). Measured at a grazing exit of 80° at the sample temperature of 40 K.

**Theoretical Details**

Calculations were performed on a slab of 13 planes (equivalent of 5 fixed and 2 relaxed $Fe_{oct}O_2$ layers), where 9 were kept fixed and 4 allowed to relax (similar to the setup used in Ref. [10]). To accurately model a metal oxide system, an effective on-site Coulomb repulsion term (Hubbard term) $U_{eff}$=3.61 [10,11] (U=4.5, J=0.89 according to Dudarev et al. [12]) was added. Slight variations of U and J are not affecting the conclusions of this work. Dispersion effects are treated by adding a non-local correlation term with a density-dependent kernel, to the base Xc functional (vdW-DF by Dion et al. [13]). Klimeš et al. [14] applied and tested the above implementation of dispersion and proposed optimized versions of the base functional of PBE to compensate small discrepancies such as imprecisions in gas phase dimer bonds, resulting in the optPBE-DF and optB88-DF [14] functionals. Dipole corrections, as implemented in VASP (IDIPOL=3 and LDIPOL=.TRUE.), according to the following Refs. [15,16], are applied.

The phonon density of states required for calculating the ZPE (zero-point energy) corrections are obtained within the harmonic approximation neglecting substrate-adsorbate interactions. Displacements are generated using

Phonopy [17]. For phonon density of state calculations, additionally, under-coordinated O$_{surface}$ atoms (labelled O*, see details in Fig. 2(A) in the main text) were also displaced, to take into account eventual changes induced by adsorbed protons.

Convergence criteria on the adsorbates and O* were further refined (down to 10$^{-9}$ eV and 0.005 eV/Å) on slabs where only the 4 topmost planes previously relaxed and optimized, remained, discarding the 9 bottom planes. With the exception of the O* atoms, the remaining 4 planes were kept fixed for the phonon calculations.

The ZPE corrected $E_{ad}$ are then given by: $E_{ad} = (E^0_{nH_2O/surf} + E^{ZPE}_{nH_2O/surf}) - (E^0_{surf} + E^{ZPE}_{surf}) - n(E^0_{H_2O} + E^{ZPE}_{H_2O})$ where $E^0_{nH_2O/surf}$, $E^0_{surf}$, $E^0_{H_2O}$ are DFT total energies of $n$ H$_2$O molecules adsorbed on the surface, the clean surface and free molecule references, respectively. $E^{ZPE}_{nH_2O/surf}$, $E^{ZPE}_{surf}$, $E^{ZPE}_{H_2O}$ are the corresponding ZPE corrections. In the case of the clean surface, only the two O* are displaced and contribute to the correction.